# Conceptional Design of the Laser Ion Source based Hadrontherapy Facility

XIE Xiu-Cui(谢修璀)[1,2]　　SONG Ming-Tao(宋明涛)[1]　　ZHANG Xiao-Hu(张小虎)[1,2]

1 Institute of Modern Physics, Chinese Academy of Sciences, Lanzhou 730000, China

2 University of Chinese Academy of Sciences, Beijing 100049, China

**Abstract:** Laser ion source (LIS), which can provide carbon beam with highly stripped state ($C^{6+}$) and high intensity (several tens mA), would significantly change the overall design of the hadrontherapy facility. A LIS based hadrontherapy facility is proposed with the advantage of short linac length, simple injection scheme and small synchrotron size. With the experience from the DPIS and HITFiL project that had conducted in IMP, a conceptional design of the LIS based hadrontherapy facility will be present with special dedication to APF type IH DTL design and simulation.

## 1 Introduction

Hadrontherapy with carbon ions is a frontier medical method in cancer therapy as well as a very important development in accelerator technology.[1][2][3] Unfortunately, due to its high cost, the hadrontherapy is still a very expensive and unaffordable medical method to most ordinary patients although it has reveal significant advantage in cure rate. So, a reduction to the cost of the hadrontherapy facility become a very meaningful work.

During the construction of an accelerator facility, ion source is a very fundamental element that determines the overall structure of the accelerator, which decides the cost of the accelerator. With an ECR (Electron Cyclotron Resonance) ion source which provide $C^{4+}$ of several hundred μA, an accumulation process in the synchrotron is needed which means multiturn injection or stripping injection scheme is necessary. Moreover, accumulation will make the beam emittance in the ring enlarged which means wider magnet gap is needed, which greatly increases the cost of the facility. Also, in order to insure a high stripping efficiency, the linac injector usually needs to be long enough to accelerate the $C^{4+}$ to a enough high energy (typically, 7MeV/u). All these reasons will make the facility complex as well as keep the cost solidly high.

LIS ion source, which can provide $C^{6+}$ of several tens mA, will solve the above problem. With its high current intensity, no accumulation in the ring is needed which means not only simplified the injection scheme but also shorten the gap of the dipole magnets which greatly reduce the costs of the facility. Furthermore, the high charge state $C^{6+}$ means no stripping process is needed which implies the linac injector could be relatively shortened.

With all these advantages, the LIS based hadrontherapy facility is quite a good option, so a conceptional design of the facility will be presented.

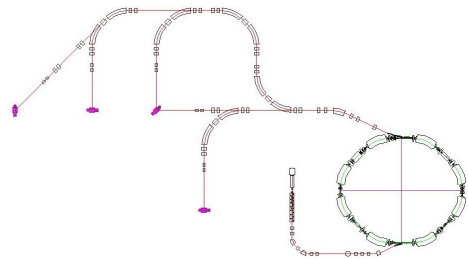

Fig 1. The layout of the LIS based hadrontherapy facility.

## 2 General considerations
### 2.1 The LIS system

The existing laser ion source in Institute of Modern Physics, Chinese Academy of Sciences (IMPCAS) uses a commercial Nd-YAG laser system and the extraction voltage is set to be 60 KV. The LIS can provide $C^{6+}$ of 21.6 mA in pulse, with laser energy of 1.437J, and the excitation time is 10.09 μs. Some of its main parameters are summarized in table 1. [4]

Table 1. The main parameters of the LIS

| Wavelength | 1064nm |
|---|---|
| Max Energy | 3J |
| Pulse Duration | 8-10ns |
| Repetition | 1Hz |
| Extraction Energy | 30KeV/u |
| Focus Length | 100mm |
| Target Dimension | 50mm×100mm×1mm |
| Max Power Density | $8.4×10^{12}W/cm^2$ |

### 2.2 The RFQ and direct plasma injection scheme (DPIS)

Since 2010, the IMPCAS has finished a series of test and commissioning to the direct plasma injection scheme. An RFQ of 100MHz, which accelerate the $C^{6+}$ from 30KeV/u to 593KeV/u, is directly joined to the extraction end of the LIS. The design injection beam is 20mA with transversal normalized RMS emittance of 0.25 π.mm.mrad, the extraction beam would be 6.5mA with transversal normalized RMS emittance of 0.35 π.mm.mrad and momentum spread of ±1.4%. The main parameters of the DPIS RFQ are listed in table 2. [5]

Table2. The main parameters of the DPIS RFQ

| Frequency | 100MHz |
|---|---|
| Cells number | 100 |
| Electrode length | 2.0m |
| Inter-vane Voltage | 120KV |
| Minimum aperture | 7.07mm |
| Modulation parameter | 1 to 2.1 |
| Synchronous phase | −90° to −20° |

### 2.3 The parameters of the DTL
### 2.3.1 The energy of the DTL

With the experience from the ready-made DPIS RFQ and the Heavy Ion Therapy Facility in Lanzhou (HITFiL) project which is now under construction, we try to give the general parameters of the DTL linac. The lattice of the synchrotron that is from the HITFiL, has the circumference of 56.173 meter and the total dipole magnet length of 25.138 meters. This means when the $C^{6+}$ is accelerated to 400MeV/u, corresponding to magnetic rigidity of 6.3654T.m, the magnetic field of the dipole magnet is working on its highest end of 1.6T. Similarly, the energy of the injection beam of the synchrotron, which is equal to the extraction energy of the DTL, is decided by the lowest end of the dipole magnet. Furthermore, the fully stripped $C^{6+}$ from the LIS makes the stripping process no longer necessary, so theoretically speaking, the extraction energy of the DTL could be as low as 2 MeV/u, but when we concern about the ripples of the power source, we want the extraction energy to be 4MeV/u, which corresponding to dipole magnetic field of 0.144T.

### 2.3.2 The Intensity of the DTL

The 4MeV/u $C^{6+}$, which corresponding to relativity β=0.0924, will circle the ring in 2.03μs. If we can keep the intensity of the extraction beam from the DTL more than 5mA, there are $1.057×10^{10}$ ions being injected into the ring in a single turn. As we consider the injection efficiency of 40%, capture efficiency of 80% and acceleration efficiency of 80%, the particle per pulse in the ring could easily achieve $2×10^9$.

### 2.3.3 The emittance of the DTL

According to the study, the growth of the normalized RMS emittance is inevitable during the acceleration. We are trying to suppress the growth of the emittance in order to reduce the gap of the magnet. Consider the transversal injection emittance of the DTL is 0.35 π.mm.mrad, we expect to control the extraction emittance under 0.5 π.mm.mrad. Also, we try to reduce the momentum spread from ±1.4% to ±0.5%.

After comparing the Alternating Phase Focusing (APF) and the Combined Zero Degree Structure (KONUS), we decide to use APF because of its simplicity in structure which makes it easy to be constructed and operated. Furthermore, the APF attracts us by its high acceleration gradient as well as enough transmission efficiency in our intensity range.

### 2.4 The modification of the synchrotron

The basic lattice structure of the synchrotron is the same as the HITFiL, we make some modification to meet the demand of LIS based scheme. Because the emittance of the injection beam is only 0.5π.mm.mrad, that a 70mm×50mm magnet aperture is enough. The reduction of the aperture would exponentially reduce the cost of the magnet, which means the primary goal of our design scheme will achieve. Moreover, the stripping injection will be replaced by the single turn injection, which remarkably simplify the injection system and cut the cost.

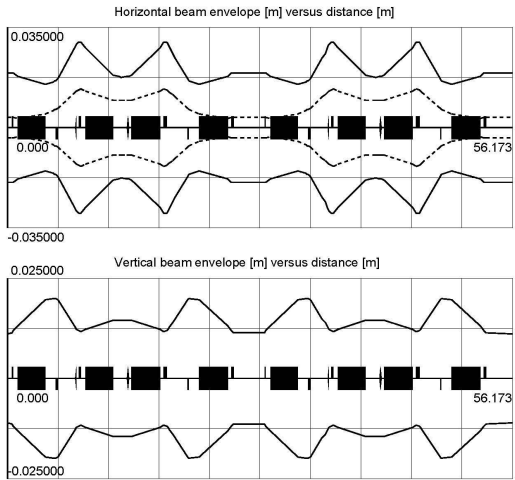

**Fig 2. The envelope of the synchrotron.**

## 3 Conceptional design of DTL linac

### 3.1 The aim of the DTL design

The aim of our DTL design is based on the following consideration. First is acceleration efficiency. The acceleration efficiency, also called the acceleration gradient, decide the length of the DTL, which will decide the cost of the accelerator-- as we have noticed, is one of our primary consideration. Secondly, the transmission efficiency. As revealed above, we hope that more than 80% transmission efficiency would be achieved. Another aim is the emittance suppression, which decides the envelope of the beam as well as the aperture of the synchrotron.

### 3.2 The principle of the DTL design

When the particles pass through the RF gap, they will "feel" electromagnetic force. According to Laplace's equation,

$$\frac{\partial^2 V}{\partial^2 x} + \frac{\partial^2 V}{\partial^2 y} + \frac{\partial^2 V}{\partial^2 z} = 0 \qquad (1)$$

Where V is the potential and x, y, z represent three orthogonal axes.

If the synchronous phase is negative, it will be focused in the longitudinal direction, which satisfy the stable acceleration condition. But at the same time, the longitudinal focusing field will bring the particles transversal defocusing effect. [6]

$$\Delta(\gamma\beta r') = -\frac{\pi q E_0 TL \sin\phi}{mc^2 \gamma_s^2 \beta_s^2 \lambda} r \qquad (2)$$

Where β,γ are the relativity factor, q and $mc^2$ are the charge and mass of the particle, $E_0TL$ is energy gain when the particle passing through the RF gap, Φ and λ are synchronous phase and wavelength, respectively, r represent any of the three axes x, y or z, and the subscript s means synchronous particle.

Conventionally, this transversal defocusing force is compensated by quadrupole magnets which obviously reduce the acceleration gradient. Whereas, the APF theory provide another solution, that is using positive synchronous phase to provide transversal focusing force. Since the beam dynamics of the APF is exclusively decided by its synchronous phase arrangement, the setting of the phase array would be crucial. In the history, several different types of APF have been designed and constructed. [7][8][9] [10][11]

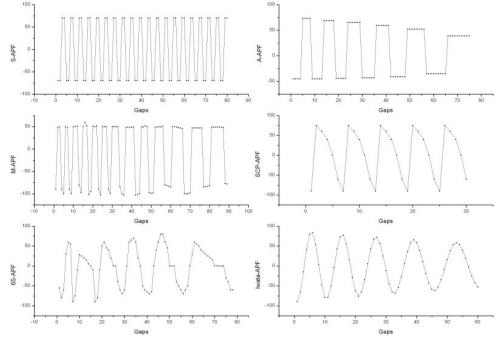

**Fig 3. Different types of APF DTL.**

Due to its relative weak focusing strength, a simple FODO type APF would be unsuitable for us. Also, as we consider the transmission efficiency as well as emittance suppression, an Iwata type APF become the best choice.

### 3.3 The DTL design

The first crucial parameter we encountered is the maximum surface electric field. As we consider the 100 MHz operation frequency, which is the same as the RFQ, the Kilpatric limit would be 113KV/cm. To avoid possible voltage breakdown, the gap voltage should be determined to insure the maximum surface field will be kept under 1.6 $E_{Kilpatric}$ which is 180KV/cm.

The phase arrangement of the Iwata type APF is described in the following function:[12]

$$\phi_s(n) = \phi_0 \cdot e^{-a \cdot n} \cdot \sin(\frac{n - n_0}{b \cdot e^{c \cdot n}}) \quad (3)$$

Where n is the cell number and $\Phi_0$, $n_0$, a, b, c are adjustable parameters used in optimization search.

With both the synchronous phase and gap voltage has been determined, the whole structure of the DTL could be established. A self-developed code is used to calculate the parameters of the structure and TRACE 3D is called to simulate the beam dynamics. At the same time, an optimization search is conducted. The result is showing in the Fig.4.

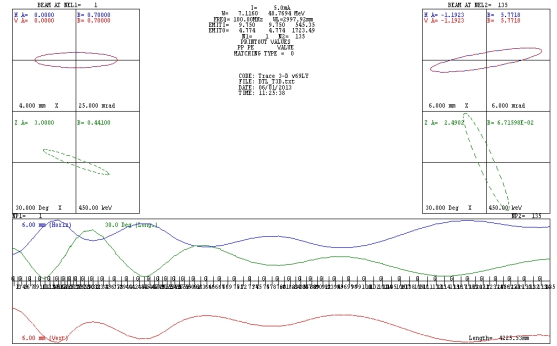

**Fig 4. The Interface of design process, with initial and terminal space phase, beam envelop and major parameters are included.**

### 3.4 The simulation

To verify our design, an end to end simulation is conduct by a particle in cell method code BEAMPATH.[13] During the simulation work, some modifications to the drift tube length is carried out to get a better transition time factor. The results are showing in the Fig. 5.

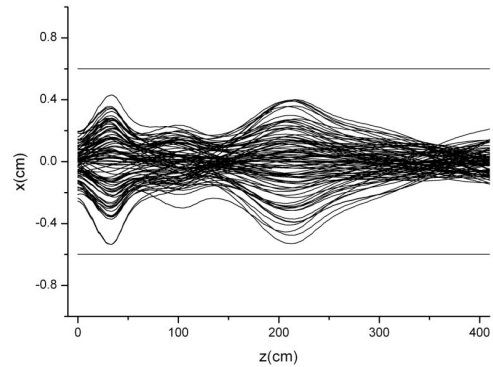

**Fig 5. The trajectory of the beam simulated by PIC method, 10000 macro-particles are used to simulate the acceleration precess. The transmission efficiency is 88.7%.**

### 4 Conclusion

As the above description, we have achieved an excellent design, which could balance the transmission efficiency, emittance suppression as well as momentum spread control. Moreover, the acceleration gradient is also quite good. With the above study we can conclude that a LIS based hadron therapy facility with high intensity, simple injection scheme, low injection energy and short linac length would be an ideal low budget choice.